\documentclass[
superscriptaddress,
 amsmath,amssymb,
 aps,
]{revtex4-2}

\usepackage{units}
\usepackage{graphicx}
\expandafter\let\csname equation*\endcsname\relax
\expandafter\let\csname endequation*\endcsname\relax

\newcommand{\comment}[1]{}

\begin{document}

\title{Transient field-resolved reflectometry at 50-100\,THz}

\author{Marcel Neuhaus}
\affiliation{Physics Department, Ludwig-Maximilians-Universit{\"a}t Munich, D-85748 Garching, Germany}
\author{Johannes Sch{\"o}tz}
\affiliation{Physics Department, Ludwig-Maximilians-Universit{\"a}t Munich, D-85748 Garching, Germany}
\affiliation{Max Planck Institute of Quantum Optics, D-85748 Garching, Germany}
\author{Mario Aulich}
\affiliation{Physics Department, Ludwig-Maximilians-Universit{\"a}t Munich, D-85748 Garching, Germany}
\affiliation{Max Planck Institute of Quantum Optics, D-85748 Garching, Germany}
\author{Anchit Srivastava}
\affiliation{Physics Department, Ludwig-Maximilians-Universit{\"a}t Munich, D-85748 Garching, Germany}
\author{D\v{z}iugas Kimbaras}
\affiliation{Physics Department, Ludwig-Maximilians-Universit{\"a}t Munich, D-85748 Garching, Germany}
\affiliation{Max Planck Institute of Quantum Optics, D-85748 Garching, Germany}
\author{Valerie Smejkal}
\affiliation{Institute for Theoretical Physics, TU Wien, A-1040, Austria}
\author{Vladimir Pervak}
\affiliation{Physics Department, Ludwig-Maximilians-Universit{\"a}t Munich, D-85748 Garching, Germany}
\author{Meshaal Alharbi}
\affiliation{Attosecond Science Laboratory, Physics and Astronomy Department, King-Saud University, Riyadh 11451, Saudi Arabia}
\author{Abdallah M. Azzeer}
\affiliation{Attosecond Science Laboratory, Physics and Astronomy Department, King-Saud University, Riyadh 11451, Saudi Arabia}
\author{Florian Libisch}
\affiliation{Institute for Theoretical Physics, TU Wien, A-1040, Austria}
\author{Christoph Lemell}
\affiliation{Institute for Theoretical Physics, TU Wien, A-1040, Austria}
\author{Joachim Burgd\"orfer}
\affiliation{Institute for Theoretical Physics, TU Wien, A-1040, Austria}
\author{Zilong Wang}
\affiliation{Physics Department, Ludwig-Maximilians-Universit{\"a}t Munich, D-85748 Garching, Germany}
\affiliation{Max Planck Institute of Quantum Optics, D-85748 Garching, Germany}
\author{Matthias F. Kling}
\email{matthias.kling@lmu.de}
\affiliation{Physics Department, Ludwig-Maximilians-Universit{\"a}t Munich, D-85748 Garching, Germany}
\affiliation{Max Planck Institute of Quantum Optics, D-85748 Garching, Germany}

\begin{abstract}
Transient field-resolved spectroscopy enables studies of ultrafast dynamics in molecules, nanostructures, or solids with sub-cycle resolution, but previous work has so far concentrated on extracting the dielectric response at frequencies below 50\,THz. Here, we implemented transient field-resolved reflectometry at 50-100\,THz (3-6\,$\mu$m) with MHz repetition rate employing 800\,nm few-cycle excitation pulses that provide sub-10\,fs temporal resolution. The capabilities of the technique are demonstrated in studies of ultrafast photorefractive changes in the semiconductors Ge and GaAs, where the high frequency range permitted to explore the resonance-free Drude response. The extended frequency range in transient field-resolved spectroscopy can further enable studies with so far inaccessible transitions, including intramolecular vibrations in a large range of systems.
\end{abstract}

\maketitle

\section{Introduction}
Mid-infrared femtosecond lasers have gained in prevalence in recent years as they open up a variety of applications in ultrafast spectroscopy, including bio-medical imaging\,\cite{Pupeza2020FRS_biosystems,Huck17}, nonlinear optics\,\cite{Keller2020OpticaWaterWindowHHG,Ghimire19}, and nanophotonics\,\cite{Guo20}. With respect to solids, such light sources permit probing charge carrier dynamics near the band edge of low-band-gap materials\,\cite{Aytac18}, carrier intraband absorptions\,\cite{steinleitner2017direct}, and the dynamics of quasiparticles, such as plasmonic or polaron excitations\,\cite{Zhong15,Cinquanta19}. Conventional pump-probe spectroscopy, either in transmission or reflection geometry, measures the change of the probe light amplitude before and after interacting with a sample. The phase information of the light field is thereby lost\,\cite{Aytac18,yeh2017ultrafastGeMIR}.

Both amplitude and phase information can be obtained with field-resolved spectroscopy\,\cite{Hwang15,Ulbricht2011RevModTHzTDSsemiconductors,Pashkin03,Huber2001NatureManyParticleScreening} which records the reflected or transmitted light electric waveform\,\cite{Cinquanta19}. Here, the time resolution is not limited by the duration of the employed pump and probe pulses. It rather depends on the duration of the pump and gating pulses used to sample the field which can both be in the visible range and have few-fs duration \cite{Huber2001NatureManyParticleScreening}.

In attosecond spectroscopy, highly-excited charge carrier dynamics in gases, nanostructures, and solids have been probed with sub-cycle time-resolution\cite{Leone16,schlaepfer2018attosecondGaAs}. Its implementation, however, required using attosecond extreme ultraviolet or x-ray pulses in a vacuum environment. This complexity has severely limited its widespread use beyond specialized laboratories, motivating the implementation of electro-optical sampling up to optical frequencies \cite{Sederberg2020NatComm}.

\begin{figure}[htbp!]
	\centering\includegraphics[width=0.95\textwidth]{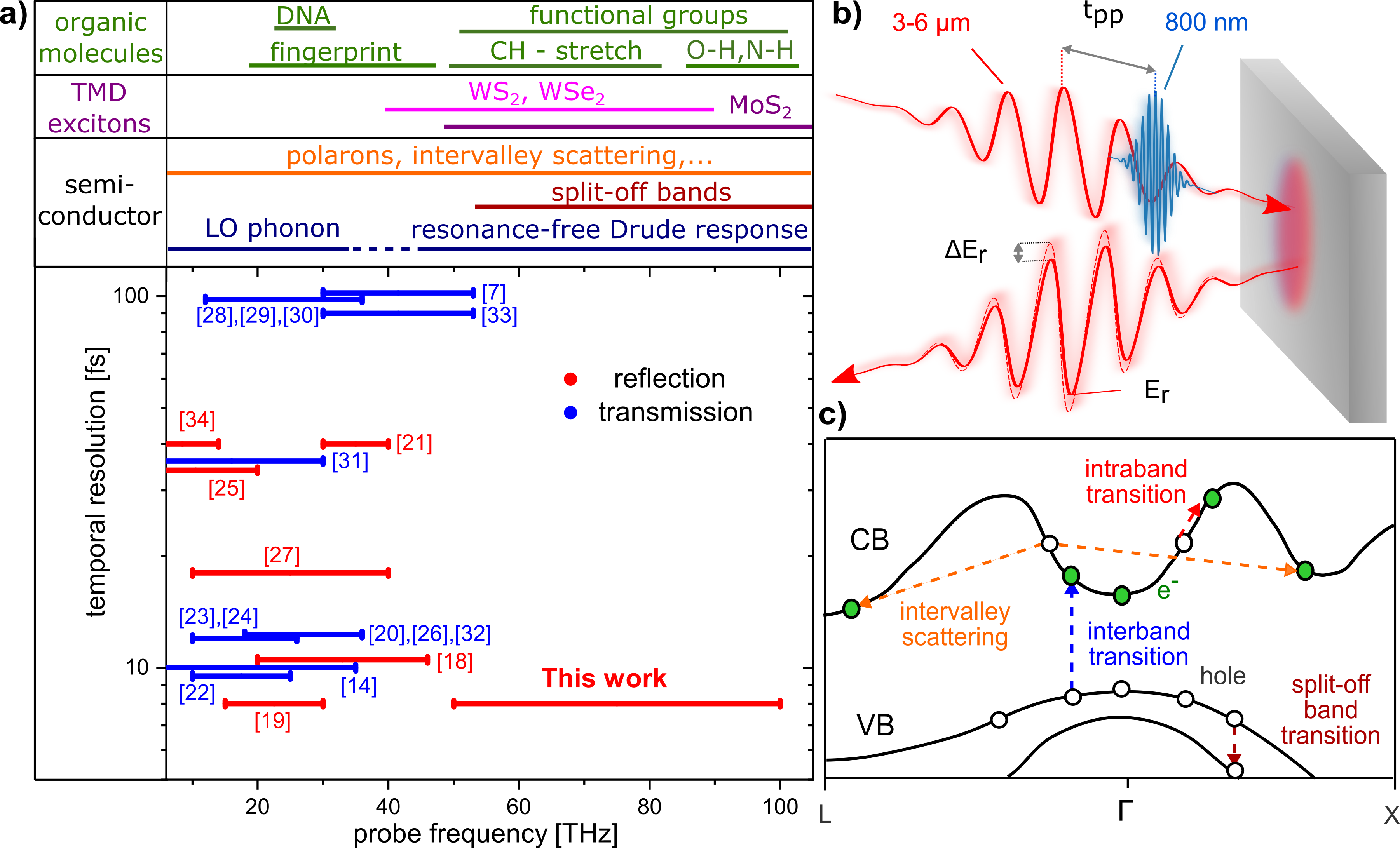}
	\caption[TFR reflectometry]{\label{Fig_scheme} (a) Overview over implementations of (transient) field-resolved spectroscopy above 20\,THz
	\cite{Eisele2014,Fischer2016,Gunter2009,Huber2016,Huber2005femtosecond,Huber2001NatureManyParticleScreening,Kim2012,Kubler2007,Lan2019,Leins2008,Mayer2014,Merkl2019,Merkl2020,Merkl2021,Otto2019,Porer2012,Poellmann2015,steinleitner2017direct,Turchinovich2017fs_ElectronMobilityGaAs}. On the top, examples of processes that can be probed in the respective frequency range are listed.(b) Scheme for TFR reflectometry in the mid-infrared. An NIR excitation pulse induces charge carrier dynamics in the solid (depicted in gray). The unfolding dynamics leads to transient changes in the reflected waveform of the time-delayed MIR pulse. (c) Schematic of laser-induced processes in semiconductors (with exemplary valleys from Ge) that can be probed with TFR reflectometry. VB: valence band, CB: conduction band.}
\end{figure}

Previous work on transient field-resolved (TFR) spectroscopy focused on extracting the dielectric response at frequencies below 50\,THz (cf. Fig.\,\ref{Fig_scheme}(a)). The frequency range above 50\,THz enables access to a range of faster processes as shown in Fig.\,\ref{Fig_scheme}(a). For example, in classical semiconductors, the faster tails of dynamics such as intervalley scattering \cite{Yu2020} and polaron excitations \cite{Wong2020}, or transitions to the split-off bands \cite{Ganikhanov1998} and the resonance-free Drude response may be investigated. In transition metal dichalcogenides (TMD) such as MoS\textsubscript{2} \cite{Cha2016,Wang2016}, WS\textsubscript{2} and WSe\textsubscript{2} \cite{Hsu2019,Poellmann2015} the internal 1s-2p transition in excitons can be probed. Most prominently, the frequency range above 50\,THz also contains the stretching modes of functional organic groups \cite{Kowlingy2019}, not just important for bio-medical applications \cite{Camp2015,WIERCIGROCH2017}, but also in molecular/organic electronics \cite{Nanova2012,Coropceanu2007}.

Here, we have developed high-repetition rate (2.1\,MHz) transient field-resolved MIR reflectometry at frequencies of 50-100\,THz (3-6\,$\mu$m) with sub-10\,fs time resolution. The capabilities of the setup are demonstrated in ultrafast light-induced charge carrier dynamics recorded for solids, characterized through the measurement of transient changes of the MIR waveform reflected from the sample surface (cf. Fig.\,\ref{Fig_scheme}(b)). The reflection geometry permits studies of optically thick and non-transparent samples including heterostructures. We performed experiments on a direct (GaAs) and an indirect (Ge) band gap semiconductor, where transient field-resolved spectroscopy allows extracting the light-induced changes of both real and imaginary parts of the materials dielectric constant. Here, the underlying charge carrier dynamics, as e.g. shown in Fig.\,\ref{Fig_scheme}(c), can be investigated. The NIR excitation pulse induces an interband transition from a valence to a conduction band. Ultrafast carrier dynamics, such as scattering and thermalization, can be monitored via changes to the reflected MIR field induced by intraband transitions.

\section{Experimental setup}
The TFR reflectometry setup is based on a home-built non-collinear optical parametric amplifier (NOPA) driven by a high-power Ytterbium-based fiber laser (Active Fiber Systems). The NOPA delivers 2\,$\mu$J sub-8\,fs pulses spanning the wavelength range 650-980\,nm at a repetition rate of 2.1\,MHz (see SI for details). These pulses are used for MIR generation via intra-pulse difference frequency generation (iDFG), transient excitation of the solid, and as gating pulses in EOS.

\begin{figure}[htbp!]
	\centering\includegraphics[width=\textwidth]{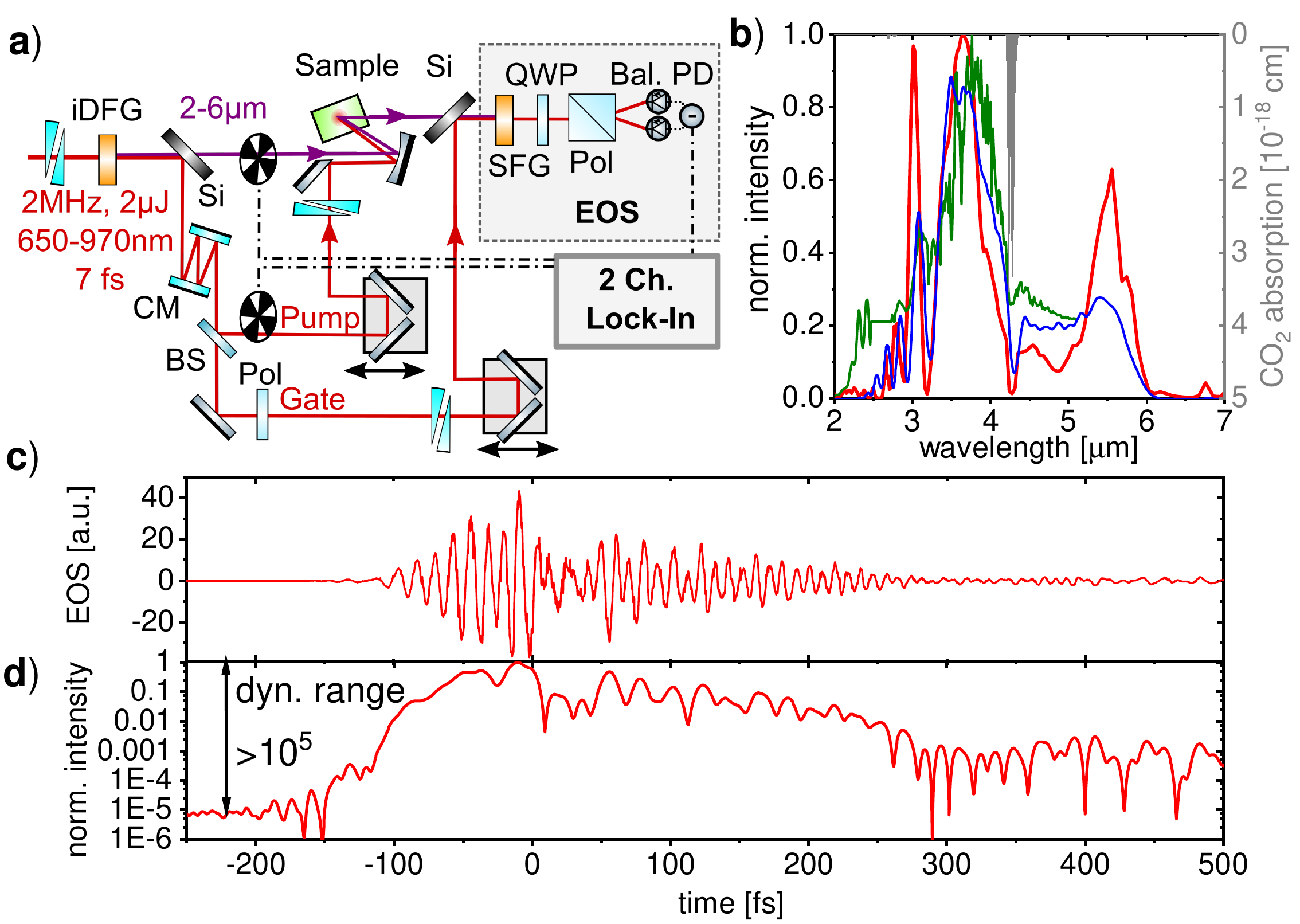}
	\caption[TFR-setup]{\label{Fig-TFR}(a) Setup for TFR-MIR reflectometry. HWP: half-wave plate, QWP: quarter-wave plate, BS: beam splitter, Pol: polarizer, Si: silicon plate, CM: chirped mirrors. (b) MIR spectrum measured with EOS (red), up-conversion spectroscopy (blue) and commercial spectrometers (green; see SI for details). The dip at 4.2\,$\mu$m is due to CO$_2$ absorption (gray). (c) Electric field recorded by EOS and (d) the temporal intensity showing 10$^5$ dynamic range with respect to the noise floor before the pulse. Signals after 300\,fs originate from the free-induction decay induced by absorption of CO$_2$ and LiIO$_3$.}
\end{figure}

Waveform-stable, octave-spanning MIR pulses are generated via iDFG in a 1\,mm thick, type-I LiIO$_3$ crystal (Fig.\ \ref{Fig-TFR}(a)) \cite{kaindl2007ultrabroadband,Kato85}. The LiIO$_3$ crystal is transparent both in the MIR region up to about 6\,$\mu$m as well as in the spectral region of the second harmonic of the driving laser. The latter is essential as otherwise two photon absorption of the driving pulses could damage the crystal. The driving NIR beam is focused to an intensity of 0.2\,TW\,cm$^{-2}$ inside the crystal. Nonlinear distortions of the beam profile were not observed. We achieved about 1\,nJ MIR pulses with more than an octave spanning bandwidth (Fig.\ \ref{Fig-TFR}(b)). To ensure the fidelity of the generated MIR light, we compare results from combining two spectrometers (InGaAs and PbSe), as well as EOS and up-conversion spectroscopy (see SI for details) \,\cite{Tidemand-Lichtenberg16,Zhu12}. The spectra measured by different methods agree qualitatively well, while both nonlinear detection methods, i.e.\ EOS and up-conversion spectroscopy, underestimate the spectral region below 3\,$\mu$m due to the phase-matching in the non-linear crystal employed in these methods. The generation and detection range with our setup could both be pushed down to 2\,$\mu$m, but at a significant loss at the long wavelengths. Since the setup is operated under ambient conditions, CO$_2$ absorption at 4.2\,$\mu$m is apparent but could be removed by purging. We note, however, that the exact spectral shape is not crucial as TFR spectroscopy detects only relative changes.

The MIR and NIR beams after the iDFG crystal are separated using a 1-mm thick silicon plate under Brewster angle for the MIR. The reflected NIR radiation is further divided into two arms by a 50:50 beam splitter (Fig.\ \ref{Fig-TFR}(a)). One arm serves as pump beam for the sample and the other serves as EOS gating pulse. Both the NIR pump and MIR probe are focused onto the sample by a spherical concave mirror under a small angle below 5$^\circ$ to the surface normal to separate the in- and outgoing beams. The reflected MIR probe beam is then re-combined with the EOS gating pulse using another silicon plate and is focused into a 0.2-mm thick LiIO$_3$ crystal for the EOS detection of the MIR electric field\,\cite{Keiber2016Electro}. The polarization rotation of the spectrally selected SFG signal (680-720\,nm) induced by the gate pulse was detected using balanced photo-diodes after a quarter wave-plate and a Wollaston prism (Fig.\ \ref{Fig-TFR}(a)). The single time domain MIR waveform is obtained by scanning the relative time delay, $t_{\mathrm{EOS}}$, between the MIR and EOS gating pulses. The pump induced MIR waveform changes are obtained by acquiring the waveform for varying pump-probe delay $t_{\mathrm{PP}}$. A typical MIR spectrum and the corresponding waveform are shown in Figs.\ \ref{Fig-TFR}(b) and (c). Comparing the field maximum with the field-free part before the pulse shows a dynamic range in detected intensity exceeding 5 orders of magnitude in the unperturbed reflected field (Fig.\ \ref{Fig-TFR}(d)).

We use a dual-modulation technique with two optical choppers to achieve the high sensitivity in the measurements. We modulate the MIR probe at a frequency of $f_0$ = 1.5\,kHz and the pump at $f_\Delta$ = 5.7\,kHz. The measured EOS signal, at each pump-probe delay, is then de-modulated using a 2-channel lock-in amplifier. The electric field of the reflected probe beam without pump, $E_r$, can be measured at the reference frequency $f_0$. At the same time, the pump-induced electric field change of the reflected MIR probe beam, $\Delta E_r$, is measured at the reference frequency $f_\Delta$. By comparing these two, the influence of drifts in the path length of the EOS interferometer as well as of laser power fluctuations on the measurement can be suppressed.
It also avoids temporal distortions of the measurement due to dispersion and limited bandwidth in the EOS crystal, discussed in Ref. \cite{Bakker1998,Beard2001}, since $E_r$ and $\Delta E_r$ are equally affected.

\section{Results and discussion}
\begin{figure}[htbp!]
	\centering\includegraphics[width=\textwidth]{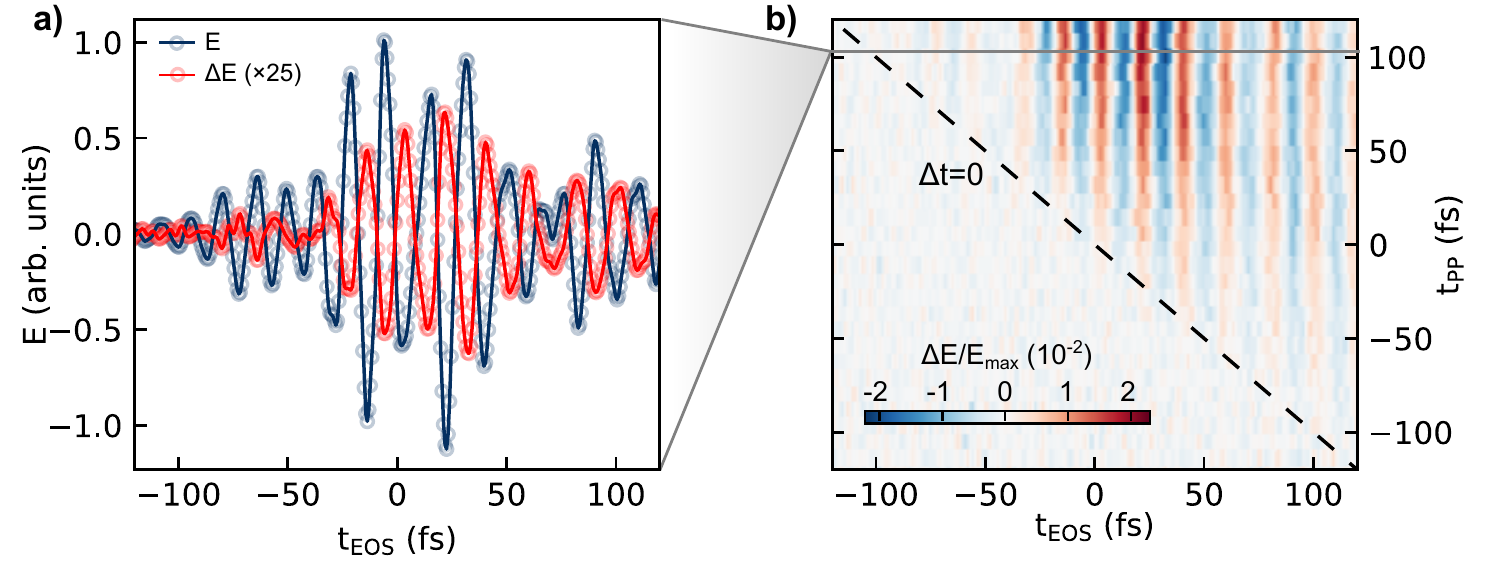}
	\caption[Experimental pump-probe signals]{\label{Fig_solid_experimental_signal} Experimental signal: (a) EOS measurement of the reflected field (blue circles) and the pump-field induced change (red circles) on GaAs for a fixed pump-probe delay of about 100\,fs (positive delays correspond to the pump arriving before the probe). The solid lines are obtained by Fourier-filtering out wavelength components below 1.2\,$\mu$m and above 14\,$\mu$m, lying outside the MIR spectrum. (b) Pump-probe delay ($t_\mathrm{PP}$) resolved relative change of the reflected field $\Delta E_\mathrm{r}/E_{\mathrm{r,max}}$, where $E_{\mathrm{r,max}}$ is the maximum field strength. The diagonal (dashed line) indicates the propagation of the pump-pulse through the EOS sampling window.}
\end{figure}

Figure\ \ref{Fig_solid_experimental_signal}(a) shows typical time domain signals of the reflected MIR field without pump ($E_\mathrm{r} (t)$, blue circles) and the pump induced change of the reflected field ($\Delta E_r (t)$, red circles) at a fixed pump-probe delay of about 100 fs from a GaAs sample. The measured $\Delta E_\mathrm{r} (t)$ is out of phase with $E_\mathrm{r} (t)$ by $\pi$, indicating a reduced reflectivity after the NIR pump excitation. Figure\,\ref{Fig_solid_experimental_signal}(b) shows a 2D plot of the transient field-resolved measurement of the change of the reflected field normalized to the maximum of the reflected field without pump, $\Delta E_\mathrm{r} (t)/E_\mathrm{r,max}$, at each pump-probe delay. The dashed diagonal line marks the simultaneous arrival time of the NIR pump and MIR probe pulses ($\Delta t_\mathrm{PP} = 0$). Changes in $\Delta E_\mathrm{r}$ appear for $\Delta t_\mathrm{PP}>0$ within tens of femtoseconds. The time resolution of the transient field-resolved experiments is only limited by the duration of the sub-10\,fs NIR pulse used as an excitation and EOS gate pulse \cite{Beard2001,Leone16,Polli2010}.

The relative change of the reflection coefficient in the frequency domain, $\Delta r(\omega)/r(\omega)$, can be calculated by Fourier transforming the measured time domain signals,
\begin{equation}
\frac{\Delta r(\omega)}{r(\omega)}=\frac{\Delta E_\mathrm{r}(\omega)}{E_\mathrm{r}(\omega)}=\left|\frac{\Delta E_\mathrm{r}(\omega)}{E_\mathrm{r}(\omega)}\right|e^{\mathrm{i}\Delta \phi (\omega)},
\end{equation}
where $r$ is the reflection coefficient for the amplitude of the electric field without pump and $\Delta \phi (\omega)$ is the pump induced phase shift in the frequency domain. Its spectral derivative, $d\Delta\phi(\omega)/d\omega$, determines the contribution to the group delay $\Delta\tau_g$ for backscattering induced by the electronic excitation of the medium. As the electronic excitation will relax and diffuse, this group delay contribution $\Delta\tau_g(t_\mathrm{PP})$ will be a function of the pump-probe delay and offers the opportunity to investigate transport and decoherence in the medium on the femto- to picosecond time scales. The relative reflectivity change of the sample, $\Delta R/R$, can be worked out from the change of the sample reflectivity $\Delta R=|r+\Delta r|^2-|r|^2$,
\begin{equation}
\frac{\Delta R}{R}=2\,\mathrm{Re}\Big(\frac{\Delta r}{r}\Big)+\mathrm{Re}\Big( \frac{\Delta r}{r} \Big)^2 + \mathrm{Im}\Big( \frac{\Delta r}{r}\Big)^2,
\end{equation}
where $\mathrm{Re}$ and $\mathrm{Im}$ are the real and imaginary parts of the complex relative change of the reflection coefficient, respectively. For small changes of the reflectivity, intensity-resolved measurements are considerably less sensitive to the imaginary part compared to the transient field-resolved approach. 
The pump-induced change of the reflectivity $\Delta r$ is linked to the refractive index $\Delta n$  through the Fresnel equations (see SI for details):

\begin{equation}
    \label{Delta_r_to_Deta_n}
	\Delta r\approx \frac{-2\Delta n}{(1+n)^2},
\end{equation}
where $n$ is the refractive index of the unexcited sample. The change of permittivity is then given by $\Delta \epsilon = (n+\Delta n)^2-n^2$. In a first approximation, after the excitation of carriers by the pump pulse, the  intraband charge carrier dynamics driven by the probe pulse may be modeled using the Drude function, assuming the measured change in the refractive index arises from free-carrier dynamics,
\begin{equation}
\label{Solid_Eq_DeltaEpsilonDrude}
\Delta \epsilon_k=\frac{-n_\mathrm{fc,k}\cdot e^2}{m_\mathrm{eff,k}\cdot\epsilon_0}\cdot\frac{1}{\omega^2-i\omega \Gamma_k},
\end{equation}
with $n_\mathrm{fc}$ the free-carrier density, $e$ the electron charge, $m_\mathrm{eff}$ the charge carrier effective mass, $\Gamma$ the Drude scattering rate (i.e.\ momentum relaxation rate) in the corresponding band, and $\epsilon_0$ the vacuum permittivity. The index $k$ labels the band and for electrons, the valley they are moving in.  Considering that the pump induced carriers by far exceed the intrinsic ones, and neglecting band gap renormalization, the total pump-induced change of the frequency-dependent dielectric function is $\Delta\varepsilon(\omega,t_\mathrm{PP})=\sum_k \Delta\varepsilon_k$, which will be dependent on the pump-probe delay $t_\mathrm{PP}$.

\begin{figure}[htbp!]
	\centering\includegraphics[width=\textwidth]{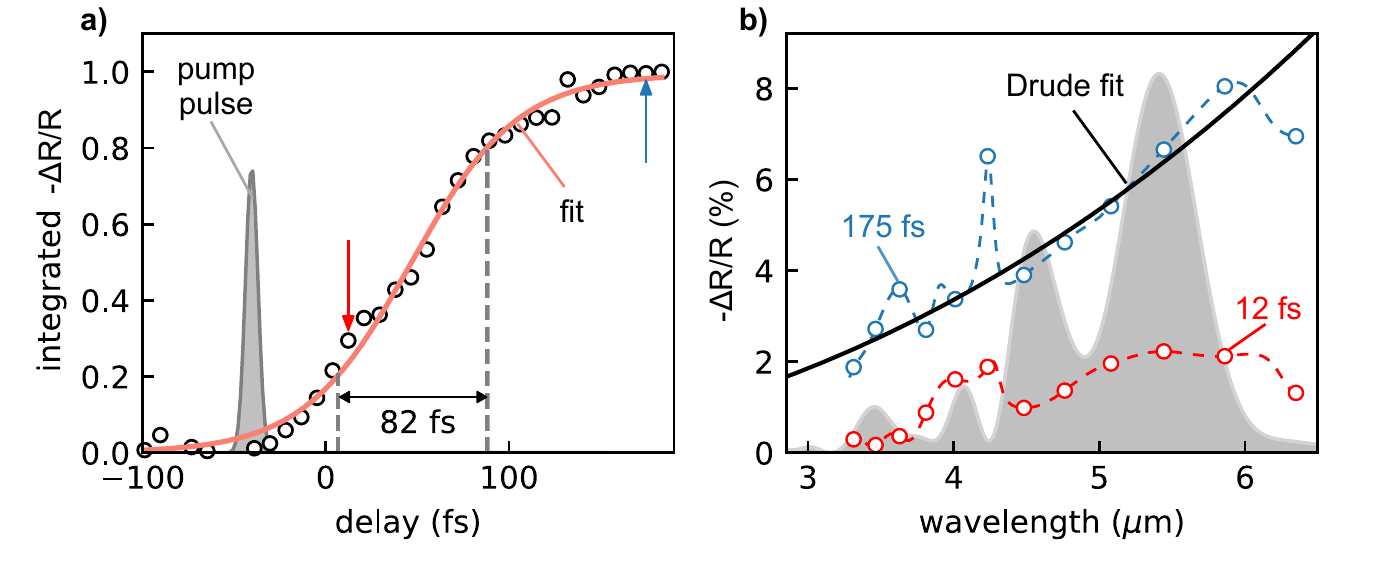}
	\caption[Femtosecond reflectivity buildup on GaAs]{\label{Fig_solid_fast_increase} Buildup of the reflectivity from GaAs: (a) Normalized real part of the experimentally measured relative reflectivity change (dots) integrated over the wavelength range 3.5-6.4\,$\mu$m and fitted to a Fermi function (red). Dashed lines indicate the rise time of 82\,fs from 20\% to 80\% of the saturation level. (b) Wavelength-resolved real part of $\Delta r/r$ for a delay of 12\,fs (red dots and dashed line) and 175\,fs (blue dots and dashed line). For the latter delay the calculated Drude expression is also shown (black solid line) which yields an excited carrier density of $1.2\cdot10^{18}\mathrm{cm}^{-3}$ and a plasma oscillation period of $\sim$60\,fs.}
\end{figure}

The spectral region of the probe light in the experiment is far away from the plasma frequency $\omega_\mathrm{pl}$ associated with the free-carrier density at the employed pump intensities. Therefore, the variation of $\Delta r(\omega)$ across the spectral window of the MIR probe is smooth and does not contain the plasmon resonance peak. More importantly, our probe frequency region is also free of phonon absorptions in semiconductors, which might affect the observed dynamics.
The pump-probe delay time $t_\mathrm{PP}$ dependence of the transient reflectivity change in GaAs integrated over the spectral window (3.5\,-\,6.4)\,$\mu$m, cf. Fig.\ \ref{Fig_solid_fast_increase}(a), displays a finite rise time of the signal with $\tau_\mathrm{rise}$\,=\,82\,fs as fitted by a Fermi function.
The delay shown here corresponds to a delay parallel to the diagonal shown in Fig. \ref{Fig_solid_experimental_signal}(b) to compensate for the different pump arrival times for different  MIR-field components \cite{Beard2000}.
A closer look at the spectral response of the signal (Fig.\ \ref{Fig_solid_fast_increase}(b)) at small pump delays ($t_\mathrm{PP}=12$ fs) and somewhat larger delay ($t_\mathrm{PP}=175$ fs), when the build-up is complete, reveals information on the temporal evolution of the response. It proceeds from an initially rather flat spectral response at short times to a Drude shape in the homogeneous approximation (Eq.\ \ref{Solid_Eq_DeltaEpsilonDrude}). Overall, both spectra exhibit a smooth shape, as expected, with the exception of the sharp peak at 4.2\,$\mu $m due to CO$_2$ absorption. The finite time response (Fig.\ \ref{Fig_solid_fast_increase}(a)) directly reflects the buildup of a Drude-type free-carrier response upon NIR photoexcitation with an excited carrier density of $1.2\cdot10^{18}\mathrm{cm}^{-3}$ as determined from the Drude fit (cf. Fig.\ \ref{Fig_solid_fast_increase}(b)).

The finite rise time of the Drude-type response in GaAs \cite{KLINGSHIRN1981} has previously been reported by Huber et al. \cite{Huber2001NatureManyParticleScreening} in optical-pump-terahertz-probe (OPTP) experiments by monitoring the appearance of a plasmon resonance peak at different time delays in the THz region. They interpreted the rise as the buildup of dressed particles and the formation of Coulomb correlations in the photoexcited electron-hole plasma on a timescale of $70-100$\,fs. This correlation time $\tau_\mathrm{cor}$ was found to be somewhat larger than the plasmon oscillation period $T_\mathrm{pl}$, $\tau_\mathrm{cor}\approx 1.6\, T_\mathrm{pl}$ and was theoretically explored using a non-equilibrium Green's function approach \cite{Huber2005femtosecond}. Our result for the rise time $t_\mathrm{rise}\approx 82$ fs is slightly smaller than the reported value of $\tau_\mathrm{cor}$. Invoking the theory for the time-resolved build-up of structured continua \cite{Wickenhauser05,Argenti13} with frequency dependent probabilities
\begin{equation}
P(\omega,t)=\left| 1 + (i-q)\frac{\Gamma}{2}\frac{\exp[i(\omega_{\mathrm{pl}}-i\Gamma/2-\omega)t]-1}{\omega-(\omega_{\mathrm{pl}}-i\Gamma/2)}\right|^2
\end{equation}
($q$: Fano asymmetry parameter), the rise time $t_\mathrm{rise}$ should, in the present case, be determined by the total decay width (or scattering rate $\Gamma$), $t_\mathrm{rise}=1/\Gamma$, rather than the plasmon period. Our result is, indeed, consistent with the Drude scattering time $\tau_s=85$ fs found in \cite{Huber2001NatureManyParticleScreening}. We thus interpret the results of Fig.\ \ref{Fig_solid_fast_increase} as an experimental verification of the time resolved build-up of a structured continuum involving free carriers in a solid. Previously, the build-up of Fano-type resonances was only observed for atomic helium in the gas phase \cite{Gruson16,Kaldun16}.

\begin{figure}[htbp!]
	\centering\includegraphics[width=\textwidth]{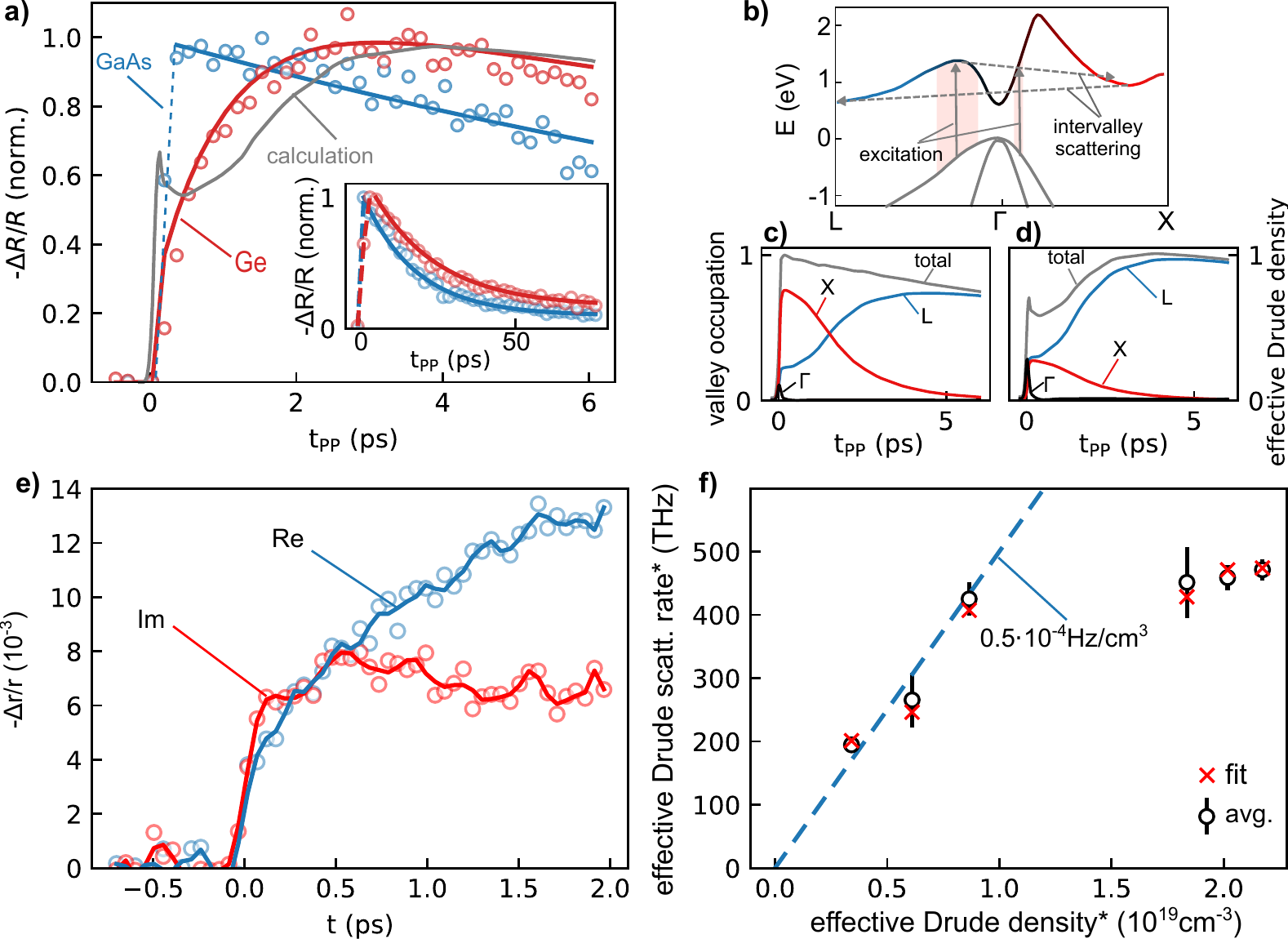}
	\caption[Intervalley scattering in Germanium]{\label{Fig_solid_Ge_intervalley} Intervalley scattering in Ge: (a) Normalized  rise in reflectivity in Ge (red dots) compared to GaAs (blue dots). The black line shows the calculated Drude reflectivity change based on the evolution of the conduction band valley occupations simulated in Ref.\ \cite{Stanton1995CalculationsFemtosecondTransmission}  for an excitation with $\hbar\omega=1.5$\,eV and the effective masses from Ref.\ \cite{dargys1994handbook} . Inset:  long-term evolution with  double-exponential fit. (b) Schematic overview of the excited electron dynamics in the band structure after photo-excitation. (c) Occupation of the individual conduction band valleys, extracted from Ref.\ \cite{Stanton1995CalculationsFemtosecondTransmission} assuming a recombination time constant of 20\,ps. (d) Individual valley contribution to the effective Drude density (carrier density divided by relative transverse effective mass) calculated from the valley occupation in (c). (e) exemplary evolution of the real (blue) and imaginary parts (red) of the reflectivity change in Ge at 5\,$\mu$m. (f) Dependence of effective Drude scattering rates on the Drude density obtained for individual wavelengths between $4.5-5.5\,\mu m$  (black bars) and from a Drude fit (red crosses).}
\end{figure}

Different from GaAs, we observe the build-up of the changes in the reflectivity of Ge on a much longer time scale extending to $\sim 3$ ps (Fig.\ \ref{Fig_solid_Ge_intervalley}(a)).
This can be understood in terms of the different band structures and, subsequently, the different charge carrier dynamics of the two samples. GaAs has a direct band gap where the NIR pump, in the range of (1.2\,$-$\,1.9)\,eV, excites charge carriers from the top of the valence band into the bottom of the conduction band around the $\Gamma$-point. The photo-induced charge carriers are expected to scatter and to relax within this $\Gamma$-valley before recombining with the hole since other valleys, e.g.\ the L- and X-valleys lie at much higher band energies.

By contrast, Ge has an indirect band gap and the interband transition occurs away from the high symmetry valleys (Fig.\ \ref{Fig_solid_Ge_intervalley}(b)). Intervalley scattering of photo-excited electrons in the conduction band plays therefore a major role in the carrier dynamics on the femto- to picosecond time scale. Indeed, Monte-Carlo simulations \cite{Stanton1995CalculationsFemtosecondTransmission, dargys1994handbook} reveal the contributions of scattering between $\Gamma$-, X- and L-valleys to the build-up of the reflectivity. The long rise time of the signal corresponds mostly to the conduction band filling of the L-valley within $(4-5)$\,ps.

We could reproduce the rise of transient reflectivity by inserting the extracted time-dependent conduction band valley populations (Fig.\ \ref{Fig_solid_Ge_intervalley}(c)) with their transverse effective masses (see SI) into a sum of Drude terms (Eq.\ \ref{Solid_Eq_DeltaEpsilonDrude}) of the $\Gamma$-, X- and L-valleys (Fig.\ \ref{Fig_solid_Ge_intervalley}(d)). Results agree qualitatively well with measured traces, except for the sharp peak at around 50 fs which results from the strong contribution of $\Gamma$-valley scattering. Overall, the different buildup behavior observed for GaAs and Ge can well be explained using the Drude-model when accounting for the different electrons' effective masses in different valleys. On much longer time scales, the relaxation processes in both GaAs and Ge behave identically, exhibiting a typical time constant of 20 ps for the data fitted up to 80 ps (inset of Fig.\ \ref{Fig_solid_Ge_intervalley}(a)).

Additional interesting insights can be gained from the separate observation of the time evolution of the imaginary and real parts of the relative reflectivity change $\Delta r/r$ as shown in Fig.\ \ref{Fig_solid_Ge_intervalley}(e). The magnitude of the real part (blue open dots) keeps rising after a sharp initial increase around $t_\mathrm{pp}=0$, while the imaginary part (red open dots) quickly flattens and even starts to decrease after 500\,fs of delay. Within the Drude model (Eq.\ \ref{Solid_Eq_DeltaEpsilonDrude}) the ratio between the imaginary and real parts of the reflection amplitude can be approximated for small changes in the refractive index to leading order by
\begin{equation}
	\frac{\mathrm{Im}\big( \Delta r/r\big)}{\mathrm{Re}\big( \Delta r/r\big)}\approx\frac{\Gamma}{\omega},\label{eq5}
\end{equation}
assuming that the unperturbed refractive index, $n$, is real valued, which applies to our case. While the effective Drude density defined as the carrier density divided by the relative transverse effective mass $n_k/m_k$, and thus $\Delta r/r$, increases through intervalley scattering, the ratio in Eq.\ \ref{eq5} is initially approximately constant until the time-dependent effective scattering rate $\Gamma$ starts to decrease after about 200\,fs. At the high excitation densities of our experiment, carrier-carrier scattering is the major contribution\cite{Meng15,Sernelius91}. However, the scattering time also depends on the carrier distribution, since for carriers excited far above the band valley, carrier-phonon scattering is significantly increased\cite{Bernardi2015PhononGaAs}. The subsequently reduced Drude scattering rate arises from the decreased carrier-carrier interactions when the electrons are scattered into different valleys as well as the reduction of the carrier-phonon scattering as the excited carrier distribution thermalizes and relaxes.

The dependence of the Drude scattering on the excited carrier density and thus the carrier-carrier scattering rate can be probed by measurement of the pump-intensity dependence of $\Gamma$ for varying pump-probe delays. The extracted Drude scattering rate for different free-carrier densities, evaluated at $t_\mathrm{pp}=1.5\,\mathrm{ps}$, when most of the density is in the L-valley, is shown in Fig.\ \ref{Fig_solid_Ge_intervalley}(f). The Drude scattering rate increases linearly with the free-carrier density for densities $0.5-1\cdot 10^{19}\mathrm{cm}^{-3}$, as indicated by the blue dashed line. At higher excitation density $\Gamma$ appears to saturate which may be an indication of the reduced available final states for e-e, e-h, and h-h scattering \cite{Okamoto2015DensityDependenceEHScattering}. The high scattering rates of the order of $\sim 100$ THz, corresponding to a Drude scattering time of $\tau_s\approx10$ fs, have also important implications for ultrafast strong-field processes in solids. Most prominently, the efficient generation of high-harmonics \cite{Ghimire19} in dielectrics involves the acceleration of a coherently driven electronic wavepacket in the conduction band and its eventual recombination with the hole. Electron-phonon and electron-electron (and hole) scattering is the main source for the decoherence of this process \cite{Floss18,Floss19}. Independent determination of Drude scattering rates from the complex reflection and transmission amplitudes thus promises novel insights into such decohering processes on the femtosecond scale.

\section{Conclusions}
We have demonstrated field-resolved transient reflectometry in the short-wavelength MIR region above 50\,THz. The approach permits recording ultrafast dynamics in molecules, nanostructures, and solids with temporal resolutions reaching 8\,fs at MHz repetition rates. The temporal evolution of the change in reflected field’s amplitude and phase can be obtained simultaneously, providing access to the pump-induced real and imaginary parts of the refractive index change of the material.

We have shown the capabilities of the technique and its extended frequency range in measurements of charge carrier dynamics in the classical semiconductors Ge and GaAs. In GaAs we could observe a build-up of the resonance-free Drude response within $82$ fs. In Ge, the real and imaginary parts of the reflectivity indicate a time-dependent Drude scattering rate due to a reduction of many-body carrier-carrier interactions following intervalley scattering.

In the future, the achievable intensity (up to TW\,cm$^{-2}$) of the NIR pump pulses will also allow for nonlinear excitations in the sample. The demonstrated field-resolved transient reflectometry at frequencies of 50 - 100\,THz paves the way towards studies involving intramolecular vibrational transitions in a wide range of systems, including in molecular/organic electronics.

\section*{Funding}
Deutsche Forschungsgemeinschaft (DFG) (SFB1375, LMUexcellent); European Research Council (ERC) (FETopen PetaCOM, COFUND Multiply, COST Action CA18234); Alexander von Humboldt Stiftung; Austrian Science Fund (FWF) (W1243 ``Solids4Fun''); TU Wien (doctoral college TU-D); Max-Planck Gesellschaft (MPG) (IMPRS-APS, MPSP, Fellow Program).

\section*{Acknowledgements}
We are grateful to Ferenc Krausz for his support and for providing suitable laboratory space. We acknowledge fruitful discussions with Rupert Huber, Thomas Nubbemeyer, and Shubhadeep Biswas, and early contributions to the design of the system by Pawel Wnuk. Z.W. acknowledges support by the Alexander von Humboldt Foundation. J.S. and V.S. are grateful for support by the International Max Planck Research School on Advanced Photon Science (IMPRS-APS). D.K. and M.F.K. acknowledge support by the Max Planck School of Photonics (MPSP). M.A. and A.M.A. are grateful for support by the Researchers Supporting Project RSP-2021/152, King Saud University, Riyadh, Saudi Arabia.

\section*{Disclosures}
The authors declare no conflicts of interest.

\section*{Data Availability Statement}
The data will be provided upon reasonable request from the authors.

\end{document}